\documentclass[11pt]{article}
\usepackage{coan,epsfig}

\def\Journal#1#2#3#4{{#1} {\bf #2}, #3 (#4)}

\def\MPL{\em Mod. Phys. Lett.}

\def\PLB{{\em Phys. Lett.}  B}
\def\PRL{\em Phys. Rev. Lett.}
\def\PR{{\em Phys. Rev.}}

\def\be{\begin{equation}}
\def\ee{\end{equation}}
\def\bea{\begin{eqnarray}}
\def\eea{\end{eqnarray}}

\begin{document}
\vspace*{4cm}
\title{FIRST RESULTS FROM CLEO-c at $\bf E_{CM}=\Psi(3770)$}

\author{ THOMAS E. COAN }

\address{Department of Physics\\
Southern Methodist University \\
Dallas, TX 75275, USA\\
E-mail: coan@mail.physics.smu.edu}

\maketitle\abstracts{ The CLEO-c detector at CESR has begun to collect
large data sets of $e^+e^- \rightarrow c\bar{c}$ events in the
center-of-mass energy range $\sqrt{s}=3 - 5\,$GeV that will greatly
enhance the size of the world's data sets of events produced at charm
threshold. Preliminary results for the pseudoscalar decay constant
$f_{D^+}$ and absolute D-meson hadronic branching fractions from an
initial integrated luminosity ${\cal L}=57\,{\rm pb^{-1}}$ collected
at $\sqrt{s}=\Psi(3770)$ are described. These charm measurements are
also relevant for B physics.}

\section{Leptonic $D^+$ Decays}
Precision measurements of leptonic (and semileptonic) decays in the
charm sector are vital for determining various elements of the CKM
matrix that describes the mixing of quark flavors and generations
induced by the weak interaction. They also can also be used to relate
CKM matrix elements to mixing measurements in the $B\bar{B}$ system
and to provide important experimental checks of Lattice QCD
calculations.  Measurements of pseudoscalar decay constants like
$f_{D^+}$ are of particular interest.

The lowest order expression for the leptonic branching fraction for the
reaction $D^+\rightarrow l^+\nu$ is given by~\cite{fd_eqn}

\begin{equation}
{\cal{B}}(D\rightarrow l\nu) = {G_F^2\over 8\pi}
M_{D^+}m^2_l(1 - {m^2_l\over m^2_{D^+}})f^2_{D^+}|V_{cd}|^2\tau_{D^+}\, ,
\label{eq:lep_decay}
\end{equation}
where $G_F$ is the Fermi coupling constant, $M_D^+$ is the $D^+$ mass,
$m_l$ is the mass of the final state lepton, $f_{D^+}$ is the
parameter encapsulating the strong physics of the process, $V_{cd}$ is
the CKM matrix element encapsulating the weak physics and quantifies
the amplitude for $c$ and $d$ quark mixing, and $\tau_{D^+}$ is the
$D^+$ lifetime. A measurement of ${\cal{B}}(D^+\rightarrow \mu^+\nu)$
coupled with PDG values for $V_{cd}$ and $\tau_{D^+}$ yields a
measurement of $f_{D^+}$. CLEO concentrates on the reaction
$D^+\rightarrow \mu^+\nu$ since for the final state charged lepton
$\mu^+$ the branching fraction is relatively large and there are fewer
final state neutrinos than for the $\tau^+$ case.

\begin{figure}
\begin{center}
\psfig{figure=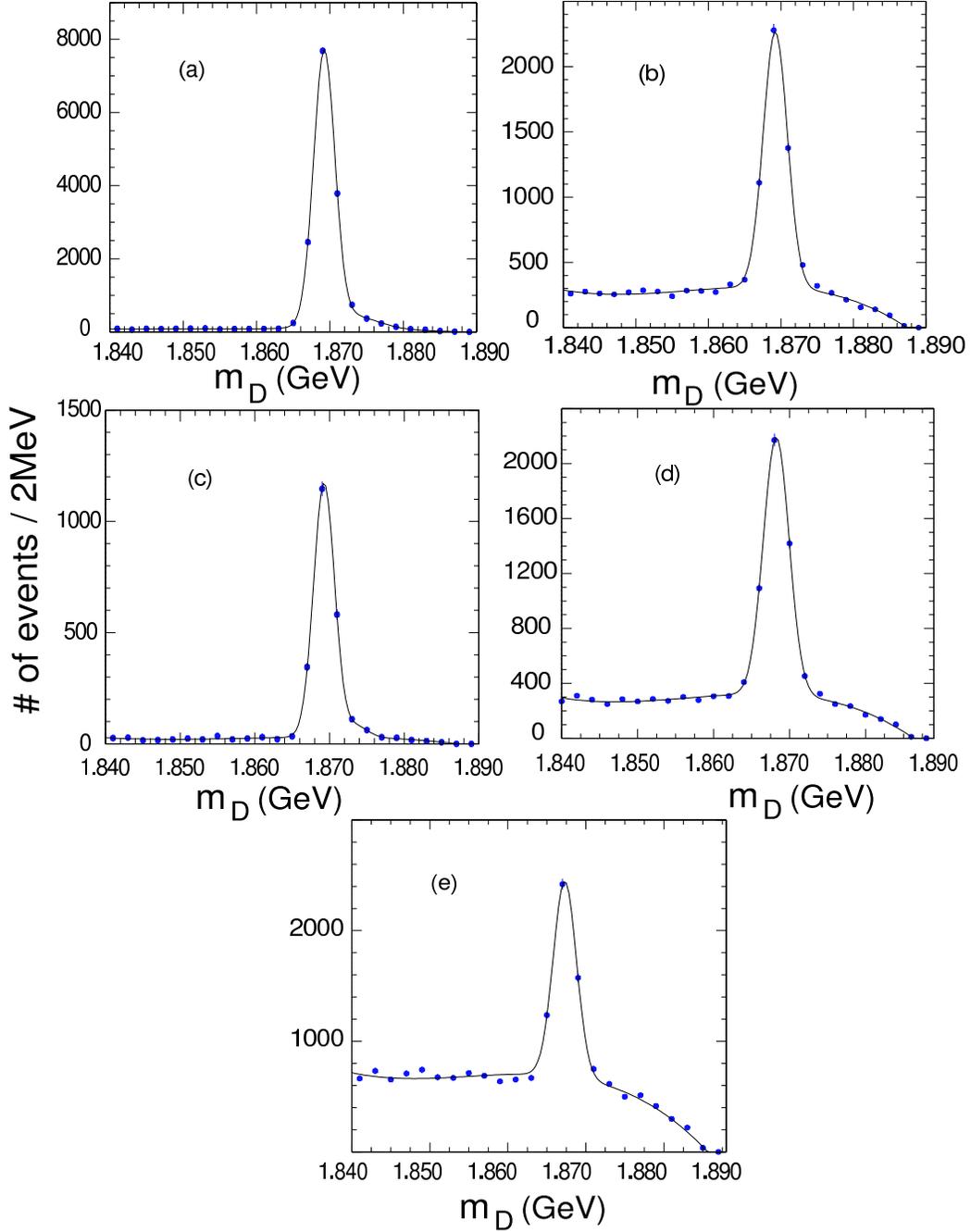,height=7.0in}
\caption{Distribution in data of the beam constrained mass $m_D$ as a
function of the $D^-$ tag mode.  The fit curves are superpositions of
Gaussian signal functions and polynomial background functions.  (a)
$D^-\rightarrow K^+\pi^-\pi^+$, (b) $D^-\rightarrow
K^+\pi^-\pi^+\pi^0$, (c) $D^-\rightarrow K^0_S\pi^-$, (d)
$D^-\rightarrow K^0_S\pi^-\pi^-\pi^+$, (e) $D^-\rightarrow
K^0_S\pi^-\pi^0$.\hfill
\label{fig:mdtags}}
\end{center}
\end{figure}

We use $57\,{\rm pb}^{-1}$ of data collected with the CLEO-c detector
at $\sqrt{s}=\Psi(3770)$.  The raw data set consists of
$D^0\bar{D^0}$, $D^+D^-$ and continuum events, with possibly a small
admixture of $\tau^+\tau^-$ and two-photon events. Events are divided
into a ``D- tag side'' and a ``signal side,'' where the tag side is
defined by attempting to fully reconstruct a candidate $D^-$ decay
into any one of the 5 modes: $D^-\rightarrow K^+\pi^-\pi^-,
K^+\pi^-\pi^-\pi^0, K^0_S\pi^-, K^0_S\pi^-\pi^-\pi^+,
K^0_S\pi^-\pi^0$. (Charge conjugate modes are implied in this analysis.)
Charged kaons and pions are identified by using both $dE/dx$
information from the central tracking chamber and information from the
ring imaging Cherenkov (RICH) counter. Neutral pions are reconstructed
using photon shower shape and location in the electromagnetic
calorimeter and $K^0_S$ particles are reconstructed from a kinematic fit
of a pair of charged pions to a displaced vertex.  Cutting on the
``beam constrained'' $D^-$ mass
$m_D=\sqrt{E^2_{beam}-(\sum\vec{p_i})^2}$, where the sum is over final
state particles, after fully reconstructing a $D^-$ candidate, is also
a powerful selection tool. The distribution in data for $M_D$ as a
function of $D^-$ tag is shown in Fig.~\ref{fig:mdtags} where the fit
curves are the superposition of Gaussian signal functions and
polynomial background functions. We find $28\,574\pm 207$ tag
candidates and $8\,765\pm 784$ background candidates.

\begin{figure}
\begin{center}
\psfig{figure=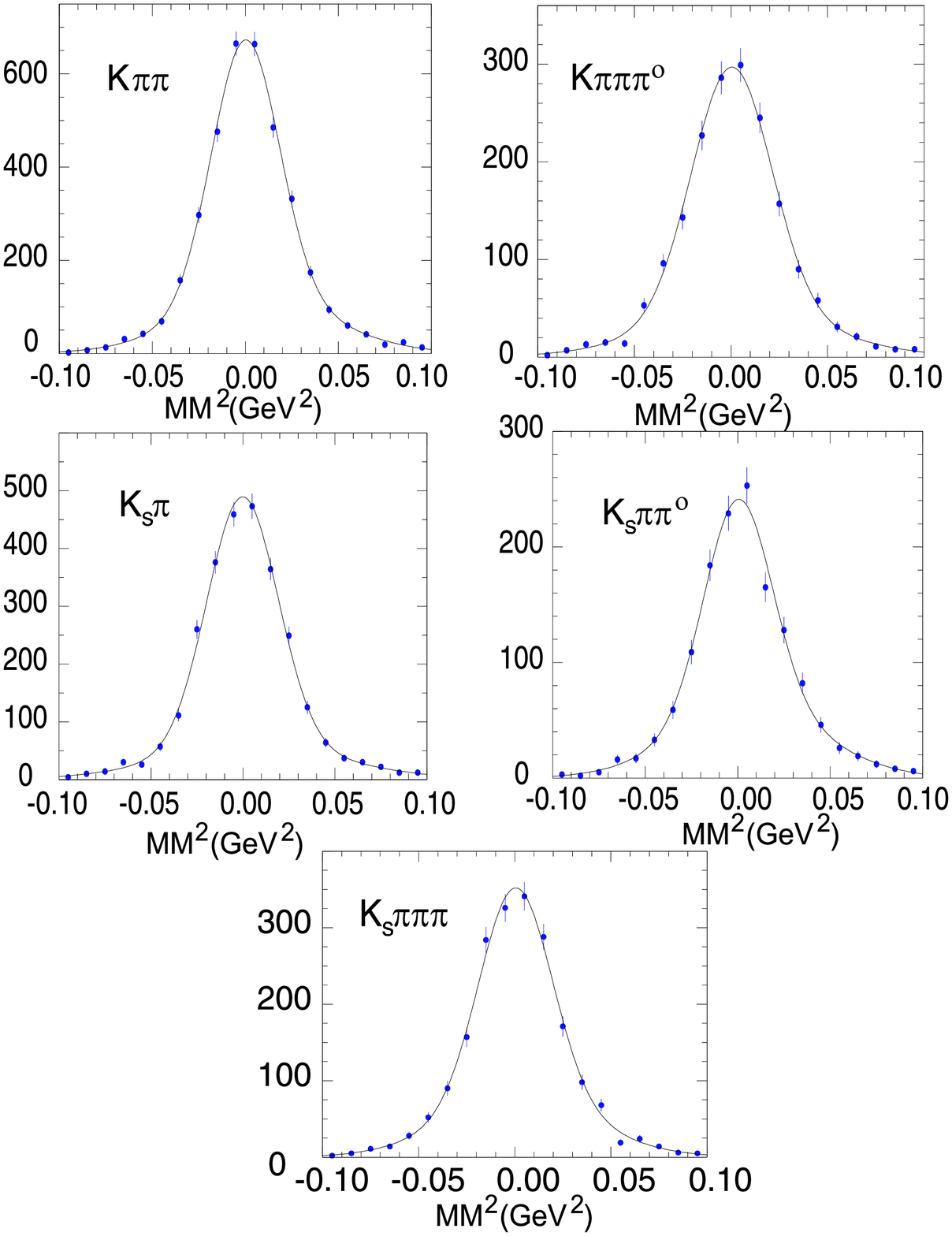,height=5.25in}
\caption{$\rm MM^2$ distribution for simulated $D^+\rightarrow \mu^+\nu$ events
as a function of the $D^-$ tag mode.\hfill
\label{fig:mm2mc}}
\end{center}
\end{figure}

Signal $D^+\rightarrow \mu^+\nu$ candidates are selected after $D^-$
tag candidates have been identified by searching for a single,
oppositely charged track, presumed to be a muon. The presence of a
neutrino is inferred by requiring that the measured value of the
missing mass squared ($\rm MM^2$) variable be near zero, where
$ 
{\rm MM^2} = (E_{beam} - E_{\mu^+})^2 - (-\vec{p}_{D^-} - \vec{p}_{\mu^+})^2
$ and $\vec{p}_{D^-}$ is the three momentum of the fully
reconstructed $D^-$. We also require that essentially no ($<
250\,$MeV) energy in the calorimeter be unmatched to a track.
Fig.~\ref{fig:mm2mc} shows the Monte Carlo distribution of the $\rm
MM^2$ variable for $D^+\rightarrow \mu^+\nu$ signal candidates as a
function of $D^-$ tag. The resolution is essentially independent of
tag mode and confirmed by comparing the resolution of
the $\rm MM^2$ distribution for both data and Monte Carlo decays of
the type $D^-\rightarrow K^0_S\pi^-$, where the same requirements are
used as for the $\mu^+\nu$ search but with the extra requirement of a
single detected $K^0_S$. Finally, we set our signal search window to be
$\pm 2\sigma$ in the $D^+\rightarrow \mu^+\nu$ $\rm MM^2$ data
distribution with $\sigma=0.028\,{\rm GeV^2}$.

\begin{figure}
\begin{center}
\psfig{figure=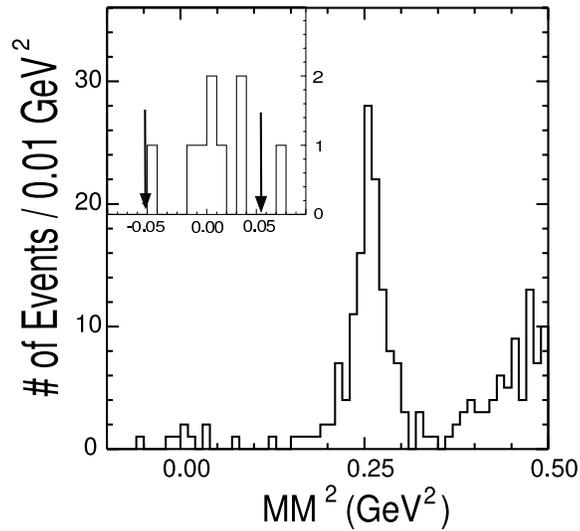,height=2.75in}
\caption{$\rm MM^2$ distribution in data using $D^-$ tags plus a
single, oppositely charged track with no unmatched energy in the
calorimeter. The insert shows 8 events inside the signal search window
denoted by the arrows.\hfill
\label{fig:mm2_sgnl}}
\end{center}
\end{figure}

Fig.~\ref{fig:mm2_sgnl} shows the data $\rm MM^2$ distribution for
$D^+\rightarrow \mu^+\nu$ candidate events after having been tagged
with a fully reconstructed $D^-$. The insert shows 8 candidates within
our search window centered about $\rm MM^2=0$. The large peak at
positive values of $\rm MM^2$ is due to $D^+\rightarrow K^0\pi^+$
events. Background events in our search window can arise from non
$D^+\rightarrow \mu^+\nu$ decay modes, misidentified $D\bar{D}$ events
and from continuum events. The likelihood of these are evaluated using
Monte Carlo simulation and the results for specific $D^+$ decay modes
are summarized in Table~\ref{tab:bkgd}. Backgrounds from
misidentified $D\bar{D}$ decays and $e^+e^-\rightarrow continuum$
events are estimated to be $0.16\pm0.16$ and $0.17\pm0.17$ events,
respectively. The total background is $1.07\pm1.07$ events, including
systematic errors.

\begin{table}[thbp]
\caption{Backgrounds from $D^+$ decay modes.\label{tab:bkgd}}
\vspace{0.2cm}
\begin{center}
\begin{tabular}{|l|c|}
\hline
\hfill Mode\hspace*{7mm}& \# of events \\ \hline
$\:D^+\rightarrow \pi^+\pi^0$&
$0.31\pm 0.04$\\
$\:D^+\rightarrow K^0\pi^+$ &
$0.06\pm0.05$\\
$\: D^+\rightarrow \tau^+\nu$&
$0.36\pm0.08$\\
$\:D^+\rightarrow \pi^0\mu^+\nu$ &
negligible\\ \hline\hline
\end{tabular}
\end{center}
\end{table}

The branching ratio ${\cal{B}}(D^+\rightarrow \mu^+\nu) =
N_{sig}/\epsilon N_{tag}$, with the number of background subtracted
signal events $N_{sig}=6.9\pm2.8$, the Monte Carlo determined single
muon detection efficiency $\epsilon=69.9\%$, and the number of $D^{\mp}$
tags $N_{tag}= 28\,574\pm207$.
\ The results are

\be
{\cal{B}}(D^+\rightarrow \mu^+\nu) = (3.5\pm1.4\pm0.6)\times 10^{-4}\;\;\; {\rm and}\;\;\;
f_{D^+} = (201 \pm 41 \pm 17)\,{\rm MeV}\,.
\label{eq:fd_result2}
\ee




\noindent These preliminary results are the first statistically
unambiguous signal for $D^+\rightarrow\mu^+\nu$ decay, consistent with
previous claims of observation~\cite{fd_bes} and with theoretical
models~\cite{fd_thy}. CLEO-c intends to collect a total integrated
luminosity ${\cal{L}}= \rm 3\,fb^{-1}$ at $\sqrt{s}=\Psi(3770)$ in the
next year.

\section{Hadronic D-meson Branching Fractions and $\bf\sigma(e^+e^-\rightarrow \Psi(3770)\rightarrow D\bar{D})$}

The absolute branching fractions of the Cabibbo allowed decays
$D^0\rightarrow K^-\pi^+$ and $D^+\rightarrow K^-\pi^+\pi^+$ (along
with $D_s^+\rightarrow\phi\pi^+$) are used to normalize the
measurements of branching fractions for nearly all other D-meson
decays. They are also used in extraction of CKM matrix elements from
$b$ and $c$ quark decay measurements.  CLEO has made preliminary
measurements of 5 important D-meson branching fractions
$D^0\rightarrow K^-\pi^+$, $D^0\rightarrow K^-\pi^+\pi^0$,
$D^0\rightarrow K^-\pi^+\pi^+\pi^-$, $D^+\rightarrow K^-\pi^+\pi^+$
and $D^+\rightarrow K_S^0\pi^+$, as well as the 2 production cross
sections $\sigma(e^+e^-\rightarrow D^0\bar{D^0})$ and
$\sigma(e^+e^-\rightarrow D^+D^-)$ at $\sqrt{s}=\Psi(3770)$. Charged
conjugate particles and decay modes are implied throughout this
analysis.

The technique for branching fraction determination follows one first
used by the Mark~III collaboration~\cite{mark3} and relies on first
fully reconstructing one of the D-mesons (``single tag'') in an event
to tag it as either $D^0\bar{D^0}$ or $D^+D^-$. Fully reconstructing
the second $D$-meson (``double tag'') then allows the absolute
branching fraction measurement of either the $D^0$ or the $D^+$
independent of the integrated luminosity or the number of $D\bar{D}$
events produced.  For instance, if $N_{D\bar{D}}$ is the total number
of $D\bar{D}$ events produced, then the number of single tag events
$N_i$ observed decaying via mode $i$ with branching fraction $B_i$ and
detected with efficiency $\epsilon_i$ is
$N_i=2N_{D\bar{D}}B_i\epsilon_i$. The number of double tag events
$N_{ij}$ with the $D$-mesons decaying via modes $i$ and $j$ is then
$N_{ij}=2N_{D\bar{D}}B_iB_j\epsilon_{ij}$ for $i\neq j$ and
$N_{ii}=2N_{D\bar{D}}B_i^2\epsilon_{ii}$ otherwise.  Finally, the
absolute branching fraction $B_i$ for a given mode $i$ is found by
taking the ratio of double tag events ($N_{ij}$) to single tag events
($N_i$). $B_i= (N_{ij}/N_j)(\epsilon_j/\epsilon_{ij})$ for $i\neq j$
and $B_i= 2(N_{ii}/N_j)(\epsilon_i/\epsilon_{ii})$ otherwise.
Including charge conjugate particles and decay modes, we use ten
single tag modes and thirteen double tag modes.

Direct use of these expressions for $B_i$ for combining errors and
measurements is problematic since $N_i$ and $N_{ij}$ are correlated,
as are measurements of $B_i$ for different tagging modes $j$. CLEO
solves this problem by using a $\chi^2$ fitting procedure that
simultaneously fits branching fractions for all $D^0$ and $D^+$
decays, and for the number of $D^0\bar{D^0}$ and $D^+D^-$ pairs
produced.  Statistical and systematic errors, backgrounds,
efficiencies and crossfeed for different modes are accounted for
directly in the fit so that experimentally measured quantities can be
accounted for in a systematic fashion.

\begin{figure}
\begin{center}
\psfig{figure=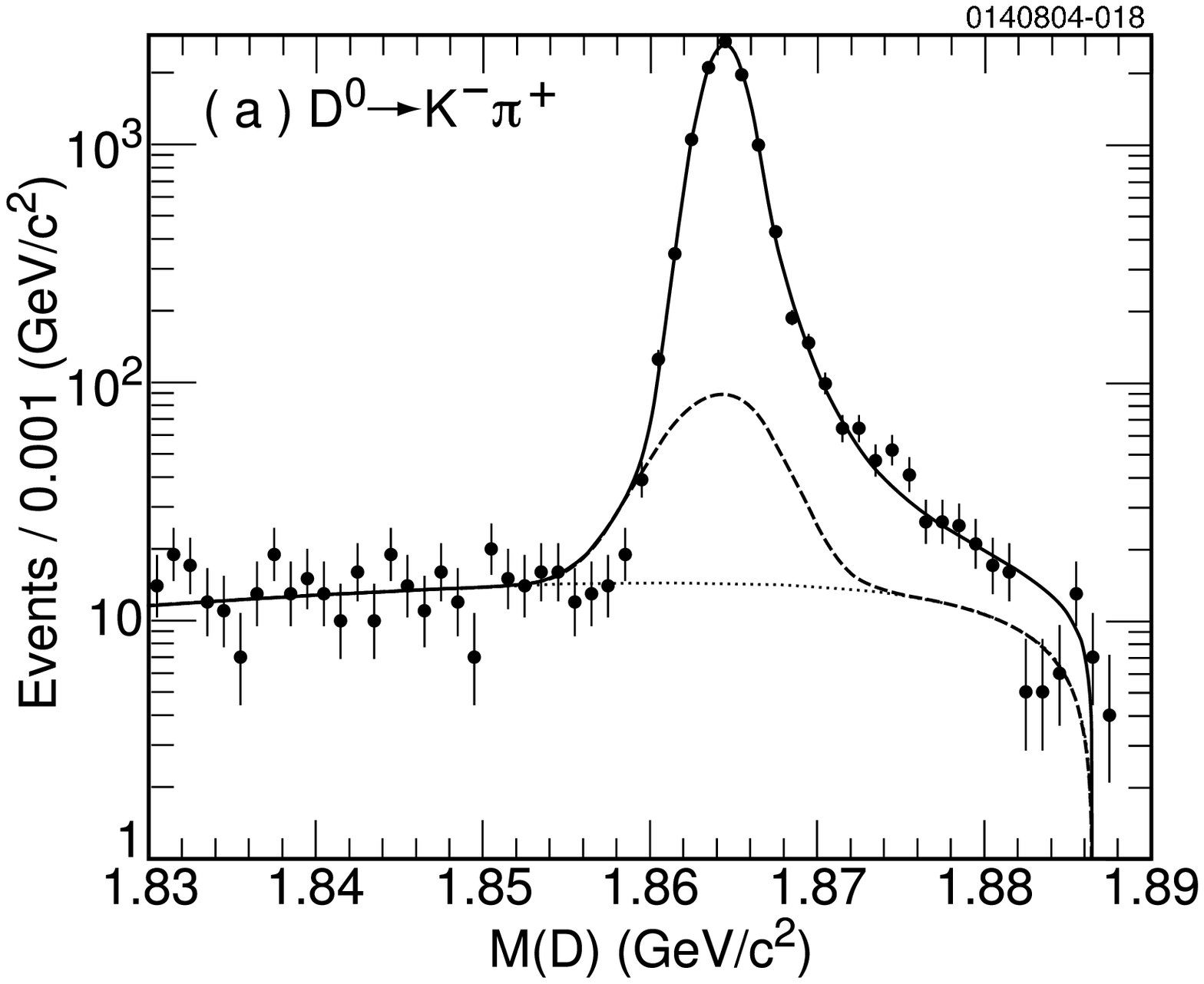,height=2.5in}
\psfig{figure=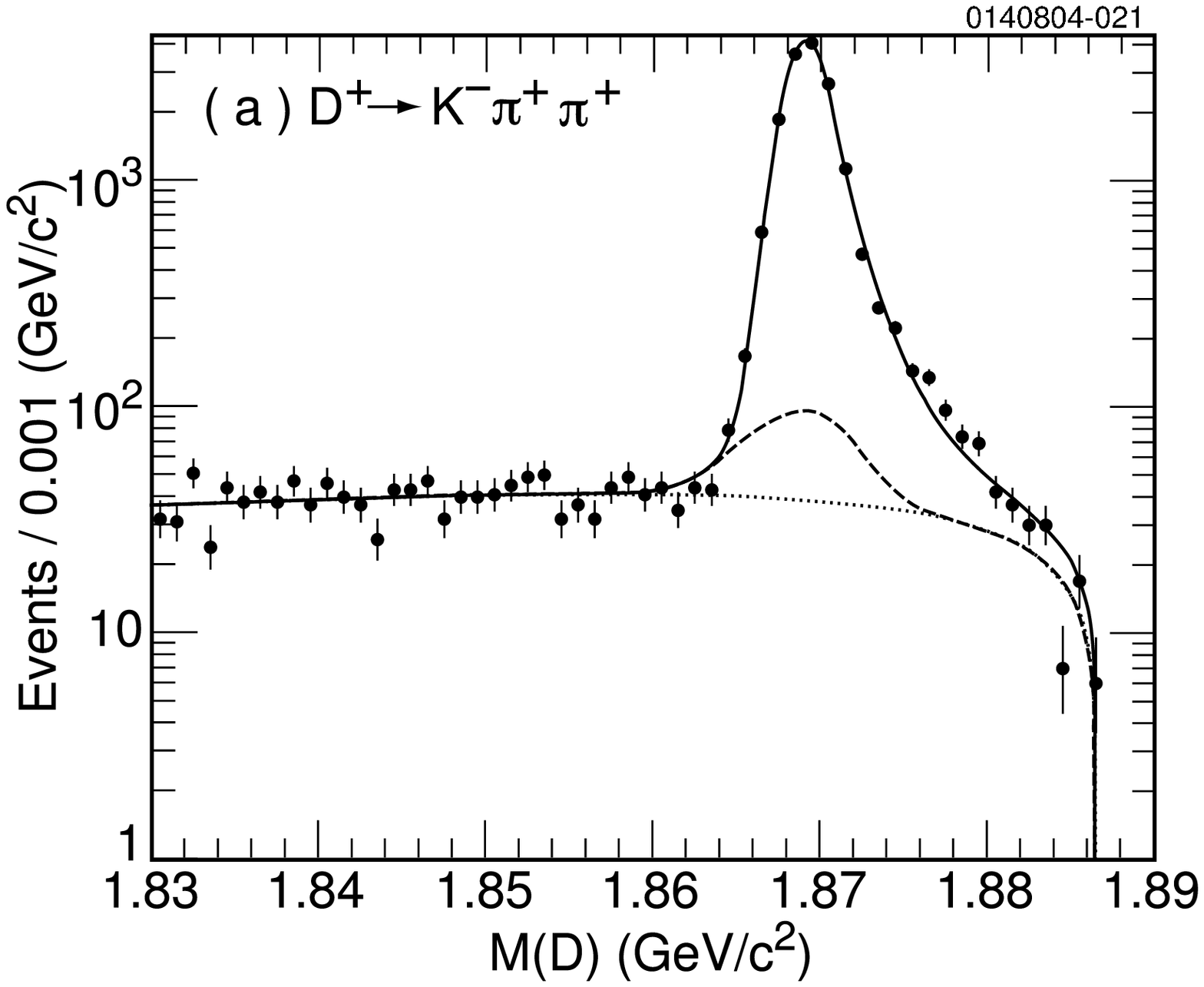,height=2.5in}
\caption{$M(D)$ distribution in data for single tag candidates for the
decay mode $D^0\rightarrow K^-\pi^+$ (left) and $D^+\rightarrow
K^-\pi^+\pi^+$ (right). The data are dots and the lines are the fit
showing the Gaussian and Crystal Ball signal functions and ARGUS background
function. \hfill
\label{fig:md1}}
\end{center}
\end{figure}

Event selection is done in a manner similar to that in the $f_{D^+}$
analysis. Key analysis variables for the final selection of $D$
candidates are the energy difference $\Delta E\equiv E(D) - E_0$,
where $E(D)$ is the total energy of the particles of the $D$ candidate
and $E_0$ is the beam energy, and the square of the $D$ candidate mass
$M^2(D)c^4 \equiv E_0^2 - p^2(D)c^2$, where $p(D)$ is the magnitude of
the $D$ candidate's three-momentum. The single tag yield in data comes
from a binned maximum likelihood fit to the $M(D)$ distribution with
line shape parameters determined from from both Monte Carlo and data.
The fit is a sum of Gaussian and Crystal Ball~\cite{cball} signal
shapes plus an ARGUS background function~\cite{argus}. $D$ and
$\bar{D}$ distributions for $M(D)$ are fit together with the same
signal parameters but independent backgrounds. For example,
Fig.~\ref{fig:md1} shows the data $M(D)$ distribution for
$D^0\rightarrow K^-\pi^+$ on the left and $D^+\rightarrow
K^-\pi^+\pi^+$ on the right.

\begin{figure}
\begin{center}
\psfig{figure=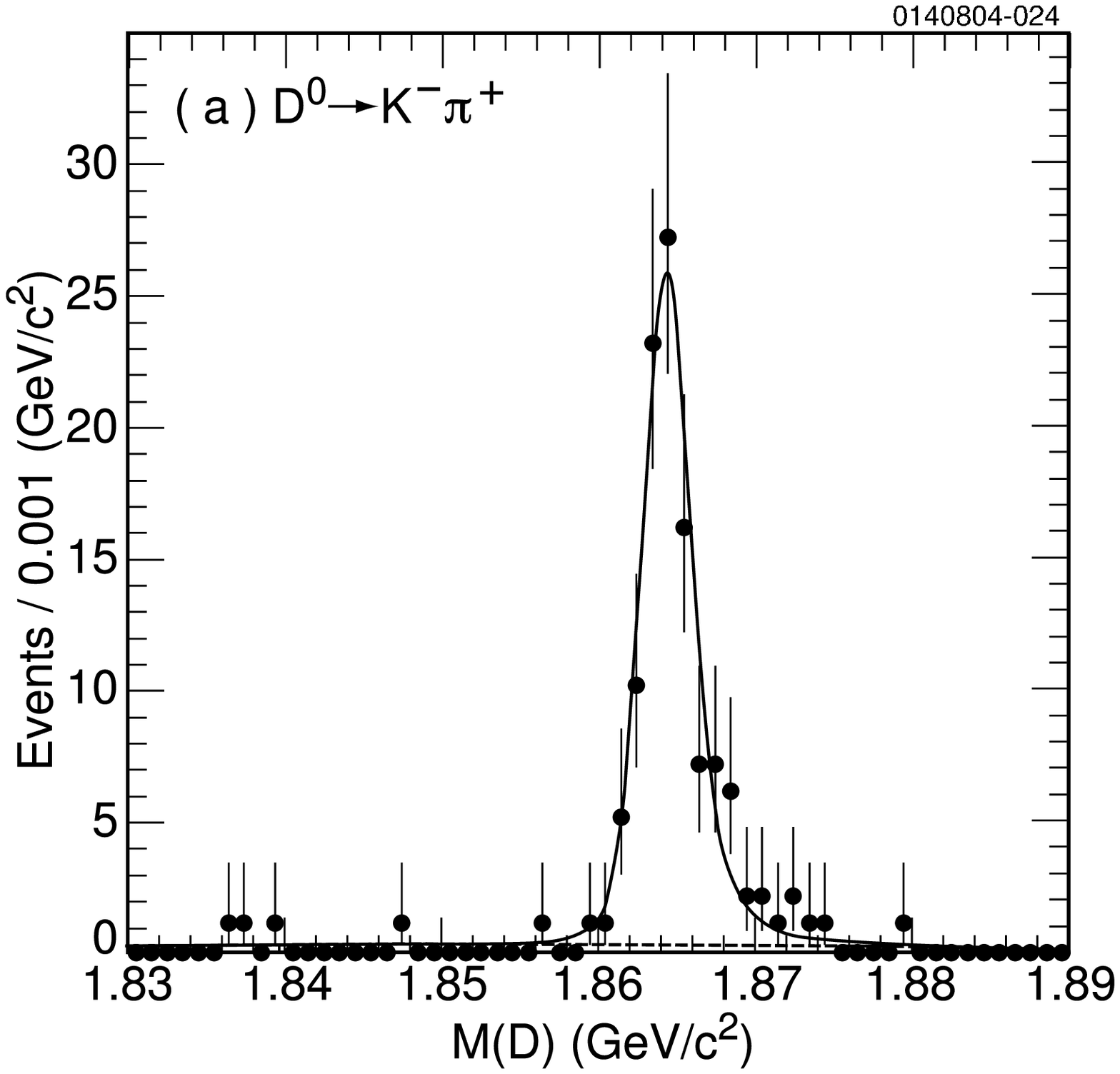,height=2.8in}
\psfig{figure=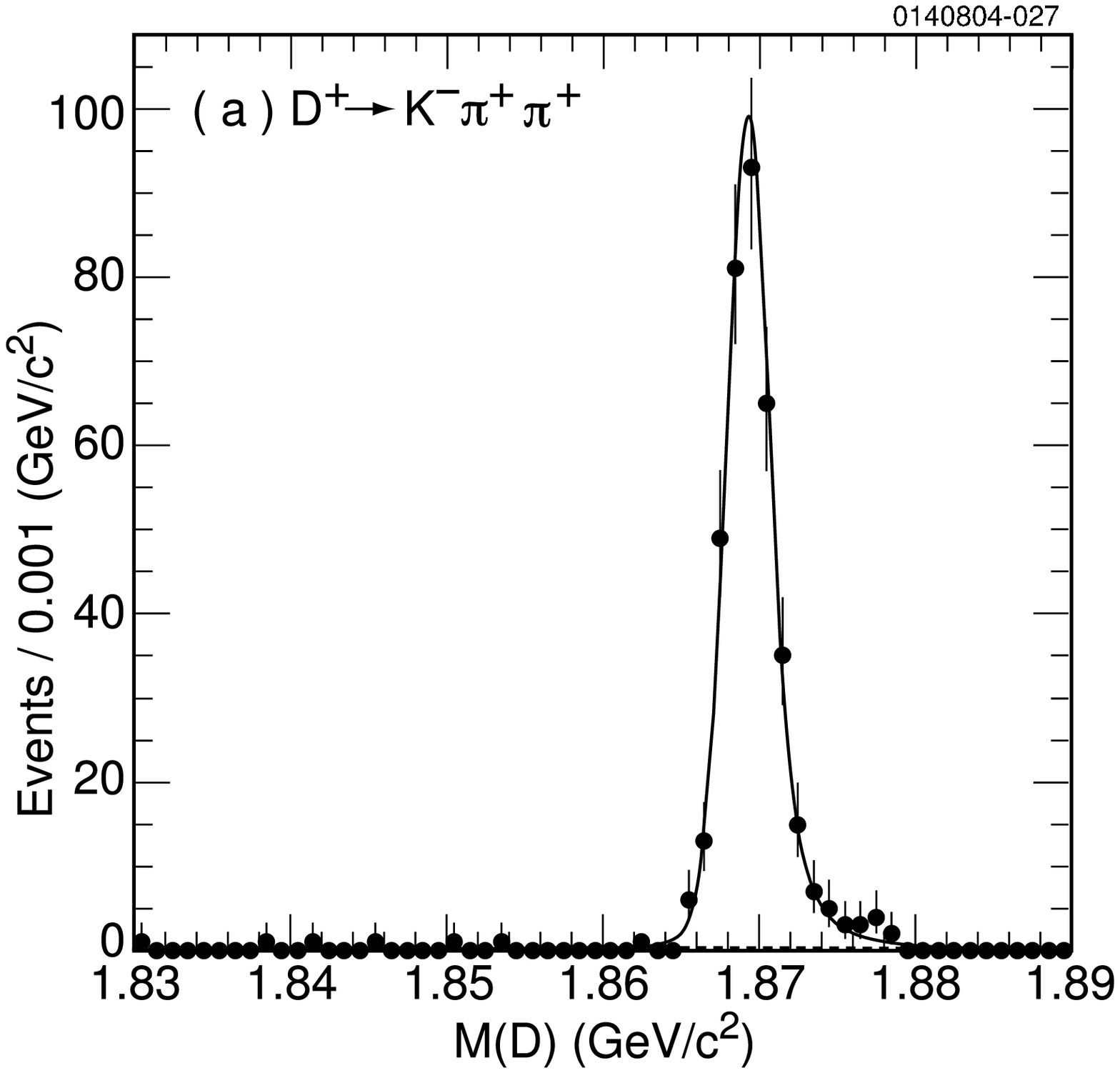,height=2.8in}
\caption{Projection of the double tag $D^0$-$\bar{D^0}$ candidate
masses onto the $M(D)$ axis for the decay mode $D^0\rightarrow
K^-\pi^+$ (and its charge conjugate mode) is shown in the left hand
plot while the right hand plot shows a similar projection of
$D^+$-$D^-$ candidate masses for the decay mode $D^+\rightarrow
K^-\pi^+\pi^+$ (and its charge conjugate mode).\hfill
\label{fig:md2}}
\end{center}
\end{figure}

The double tag yield in data is extracted from an unbinned maximum
likelihood fit to the two-dimensional distribution of
$M(D)$-$M(\bar{D})$ derived from double tag candidate events. Again,
shape parameters are determined from a combination of Monte Carlo and
data events. As an example, Fig.~\ref{fig:md2} shows the projection of
the fit onto the $M(D)$ axis for the case when the $D$-mesons are
reconstructed as $D^0\rightarrow K^-\pi^+$ and $\bar{D^0}\rightarrow
K^+\pi^-$ (left-hand side). The right-hand side shows a similar
projection for the case $D^+\rightarrow K^-\pi^+\pi^+$ and
$D^-\rightarrow K^+\pi^-\pi^-$.

The branching fraction fit for the $D$-meson branching fractions
listed above as well as $N_{D^0\bar{D^0}}$ and $N_{D^+D^-}$, uses
event yields and efficiencies for ten single tag and thirteen double
tag modes. Crossfeed and backgrounds are accounted for in the fit.
The $\chi^2$ of the fit is 8.9 for 16 degrees of freedom (confidence
level = 92$\%$) and the fit results are shown in
Table~\ref{tab:results}. The branching fractions are consistent with,
but somewhat higher than, the current PDG values. In nearly all cases,
our statistical error for the branching fractions and their ratios is
less than the statistical error of the individual measurements
comprising the PDG average. Cross sections are extracted using our
current integrated luminosity ${\cal{L}} = 57\,{\rm pb^{-1}}$. The
statistical and systematic errors of these preliminary results should
drop markedly when CLEO accumulates its anticipated $3\,{\rm fb^{-1}}$
of integrated luminosity.

\begin{table}[thbp]
\caption{Fit results for branching fractions and $D\bar{D}$ pair
yields, and inferred production cross sections. First error is statistical
and second is systematic.\hfill\label{tab:results}}
\vspace{0.2cm}
\begin{center}
\begin{tabular}{|l|l|}
\hline
\hfill Parameter\hspace*{14mm}& Fitted/Inferred Value \\ \hline
$\:{\cal{B}}(D^0\rightarrow K^-\pi^+$)&
$(3.92 \pm 0.08\pm0.23)\%$\\
$\:{\cal{B}}(D^0\rightarrow K^-\pi^+\pi^0)$&
$(14.3\pm0.3\pm1.0)\%$\\
$\: {\cal{B}}(D^0\rightarrow K^-\pi^+\pi^-\pi^+)$&
$(8.1\pm0.2\pm0.9)\%$\\
$\:{\cal{B}}(D^+\rightarrow K^-\pi^+\pi^+)$ &
$(9.8\pm0.4\pm0.8)\%$\\
$\:{\cal{B}}(D^+\rightarrow K_S^0\pi^+)$ &
$(1.61\pm0.08\pm0.15)\%$\\
$N_{D^0\bar{D^0}}$& $1.98 \times 10^5$\\
$N_{D^+D^-}$& $1.48\times 10^5$\\
\hline
$\sigma(e^+e^-\rightarrow \Psi(3770)\rightarrow D^0\bar{D^0})$ &
$3.47\pm 0.07\pm0.15\,$nb\\
$\sigma(e^+e^-\rightarrow \Psi(3770)\rightarrow D^+D^-)$ &
$2.59\pm 0.11\pm0.11\,$nb\\
$\sigma(e^+e^-\rightarrow \Psi(3770)\rightarrow D\bar{D})$ &
$6.06\pm 0.13\pm0.22\,$nb\\
\hline\hline
\end{tabular}
\end{center}
\end{table}

\section*{Acknowledgments}
The author would like to thank the conference organizers for the
invitation and the stimulating environment, the U.S. Department of
Energy for its support under contract DE-FG03-95ER40908, and W.M. Sun
for useful discussion.

\end{document}